\documentclass[aps,prb,preprint,superscriptaddress,amsmath]{revtex4-1}
\usepackage{graphicx}
\usepackage{epstopdf}
\newcommand{\wn}{\ensuremath{\mathrm{cm}^{-1}}}
\begin{document}

% Use the \preprint command to place your local institutional report
% number in the upper righthand corner of the title page in preprint mode.
% Multiple \preprint commands are allowed.
% Use the 'preprintnumbers' class option to override journal defaults
% to display numbers if necessary
%\preprint{}

%Title of paper
\title{Unexpectedly Large Tunability of Lattice Thermal Conductivity of
Monolayer Silicene via Mechanical Strain}
% repeat the \author .. \affiliation  etc. as needed
% \email, \thanks, \homepage, \altaffiliation all apply to the current
% author. Explanatory text should go in the []'s, actual e-mail
% address or url should go in the {}'s for \email and \homepage.
% Please use the appropriate macro foreach each type of information

% \affiliation command applies to all authors since the last
% \affiliation command. The \affiliation command should follow the
% other information
% \affiliation can be followed by \email, \homepage, \thanks as well.
\author{Han Xie}
%\surname{Xie}
\affiliation{University of Michigan-Shanghai Jiao Tong
 University Joint Institute, Shanghai Jiao Tong University,
 Shanghai 200240, China}
\author{Tao Ouyang}
%\surname{Ouyang}
\affiliation{Institute of Mineral Engineering, Division of
 Materials Science and Engineering, Faculty of Georesources and
 Materials Engineering, RWTH Aachen University, Aachen 52064, Germany}
\author{\'Eric Germaneau}
%\surname{Germaneau}
\affiliation{Center for High Performance Computing, Shanghai Jiao Tong
 University, Shanghai 200240, China}
\author{Guangzhao Qin}
%\surname{Qin}
\affiliation{Institute of Mineral Engineering, Division of
 Materials Science and Engineering, Faculty of Georesources and
 Materials Engineering, RWTH Aachen University, Aachen 52064, Germany}
\author{Ming Hu}
%\surname{Hu}
\email{Authors to whom correspondence should be addressed. E-mail
 addresses: hum@ghi.rwth-aachen.de and hua.bao@sjtu.edu.cn.}
\affiliation{Institute of Mineral Engineering, Division of
 Materials Science and Engineering, Faculty of Georesources and
 Materials Engineering, RWTH Aachen University, Aachen 52064, Germany}
\affiliation{Aachen Institute for Advanced Study in Computational
 Engineering Science (AICES), RWTH Aachen University, Aachen 52062,
 Germany}
\author{Hua Bao}
%\surname{Bao}
\email{Authors to whom correspondence should be addressed. E-mail
 addresses: hum@ghi.rwth-aachen.de and hua.bao@sjtu.edu.cn.}
\affiliation{University of Michigan-Shanghai Jiao Tong
 University Joint Institute, Shanghai Jiao Tong University,
 Shanghai 200240, China}
%\author{}
%\email[]{Your e-mail address}
%\homepage[]{Your web page}
%\thanks{}
%\altaffiliation{}
%\affiliation{}

%Collaboration name if desired (requires use of superscriptaddress
%option in \documentclass). \noaffiliation is required (may also be
%used with the \author command).
%\collaboration can be followed by \email, \homepage, \thanks as well.
%\collaboration{}
%\noaffiliation

\date{\today}

\begin{abstract}
Strain engineering is one of the most promising and effective routes
toward continuously tuning the electronic and optic properties of
materials, while thermal properties are generally believed to be
insensitive to mechanical strain. In this paper, the
strain-dependent thermal conductivity of monolayer
silicene under uniform bi-axial tension is computed by solving the phonon
Boltzmann transport equation with force constants extracted from
first-principles calculations. Unlike the commonly believed
understanding that thermal conductivity only slightly decreases with increased
tensile strain for bulk materials, it is found that the thermal conductivity of
silicene first increases dramatically with strain and then slightly
decreases when the applied strain increases further. At a tensile strain of 4\%, the highest
thermal conductivity is found to be about 7.5 times that of unstrained one. Such an unusual
strain dependence is mainly attributed to the dramatic enhancement
in the acoustic phonon lifetime. Such enhancement
plausibly originates from the flattening of the buckling of the
silicene structure upon stretching, which is unique for silicene as
compared with other common two-dimensional materials. Our findings offer perspectives of modulating
the thermal properties of low-dimensional structures for applications
such as thermoelectrics, thermal circuits, and nanoelectronics.
\end{abstract}

% insert suggested PACS numbers in braces on next line
% insert suggested keywords - APS authors don't need to do this
%\keywords{}

%\maketitle must follow title, authors, abstract, \pacs, and \keywords
\maketitle

% body of paper here - Use proper section commands
% References should be done using the \cite, \ref, and \label commands
\section{Introduction}
% Put \label in argument of \section for cross-referencing
%\section{\label{}}
Two-dimensional (2D) materials have been extensively studied in the
past decade because of their novel physical and chemical properties
\cite{novoselov_electric_2004,seol_two-dimensional_2010,yan_thermal_2014}
and potential applications.\cite{renteria_graphene_2014,tao_silicene_2015} For
example, it has been found that graphene has extremely high thermal
conductivity,\cite{balandin_superior_2008} which has great potential
in the applications including electronic cooling and composite
materials. Silicene is the silicon counterpart of graphene and another
typical 2D material with a honey-comb lattice structure. Compared to
graphene, silicene is more compatible with silicon-based
semiconductor devices and therefore has greater potential in
nanoelectronic applications. Silicene has also been found to have
opened a tunable band gap when a transverse electric field is
applied.\cite{drummond_electrically_2012,ni_tunable_2012,kaloni_hole_2013}
Monolayer silicene has been successfully fabricated on substrates such as
Ag(110),\cite{aufray_graphene-like_2010}
Ir(111),\cite{meng_buckled_2013} and
Ag(111)\cite{lalmi_epitaxial_2010}
surfaces. Recently, Tao \emph{et al.} have demonstrated silicene
transistors operating at room temperature.\cite{tao_silicene_2015}
Although the performance is still moderate and the lifetime of this
transistor is only a few minutes, it has attracted significant research
interest in silicene based
devices.\cite{nguyen_silicene_2014,lian_effects_2015,schwierz_two-dimensional_2015}

On the other hand, the intrinsic physical properties of silicene, such
as lattice thermal conductivity, have been an active area of research.
Although the thermal conductivity of silicene has not been measured in
experiments due to the difficulty of synthesizing free standing
silicene, several numerical simulations have predicted the thermal
conductivity of silicene and the results at 300K range from 5 to
69~W/mK.\cite{li_vacancy-defectinduced_2012,ng_molecular_2013,pei_tuning_2013,hu_anomalous_2013,xie_thermal_2014,gu_first-principles_2015}
Most of the numerical simulations are based on classical molecular
dynamics and the discrepancy of results mainly arises from the different
interatomic interaction potentials used. Notably,
first-principles-based lattice dynamics predicted that the thermal
conductivity of silicene is in the range of 20-30~W/mK,
\cite{xie_thermal_2014,gu_first-principles_2015} which should be more
reliable. In our previous first-principles
calculations,\cite{xie_thermal_2014} the thermal conductivity of
9.4~W/mK at 300~K was not refined due to the small cutoff used for the anharmonic force constant calculation and not imposing the acoustic sum rule.\cite{li_thermal_2012} Our new calculation with larger
cutoff and denser $q$-mesh gives a value of around 25~W/mK, which is
consistent with other literature.\cite{gu_first-principles_2015} Despite recent efforts to describe the
properties of unstrained silicene, in real applications, nanoscale
devices usually contain residual strain after fabrication.
\cite{bhowmick_effect_2006} It is thus important to investigate
possible strain effects on the property of silicene. It was found that a
mechanical tensile strain less than 5\% could tune the electronic
structure of silicene\cite{yan_tuning_2015} and larger tensile strain
(7.5\%) could induce a semimetal-metal
transition.\cite{liu_strain-induced_2012} On the other hand, using
first-principles it has been demonstrated that the silicene structure
remains buckled even when 12.5\% tensile strain is
applied.\cite{liu_strain-induced_2012,lian_strain_2013}

In comparison to the structural and electronic properties, the strain
effect on the lattice thermal conductivity of silicene is less
investigated. Pei \emph{et al.}\cite{pei_tuning_2013} and Hu \emph{et al.}
\cite{hu_anomalous_2013} investigated the effect of uniaxial
strain on the thermal conductivity based on the classical
non-equilibrium molecular dynamics method. Pei \emph{et al.}
studied tensile strain up to 12\% and concluded that the thermal
conductivity first increases slightly (around 10\% increment) and then
decreases with an increased amount of tensile strain. Hu \emph{et al.}
found that the thermal conductivity of silicene sheet and silicene
nanoribbon experiences monotonic increase by a factor of 2 with tensile
strain up to 18\%. The modified embedded-atom method
(MEAM)\cite{baskes_modified_1992} and original Tersoff
potential\cite{tersoff_modeling_1989} were used in their simulations,
respectively. However, both potentials are developed for bulk silicon,
so directly applying those potentials to the new 2D silicene
structure is questionable. For example, the Tersoff potential cannot
even reproduce the buckled structure of silicene and the MEAM potential
seems to overestimate the buckling distance. It is well known that the
interatomic potential directly determines the quality of classical
molecular dynamics simulation. Therefore, in order to precisely predict
the strain effect on the lattice thermal conductivity of silicene and
identify the underlying mechanism, it is necessary to calculate the
lattice thermal conductivity of silicene under different strains using
a more accurate method.

In this paper, the strain dependent thermal conductivity of monolayer
silicene is calculated based on single mode relaxation time
approximation (RTA) and iterative solution of the Boltzmann Transport
Equation (BTE), where the harmonic and anharmonic force constants are
determined using first-principles calculations. The contributions of
different modes under different strains are analyzed. The governing
mechanisms are analyzed and compared with other materials.

\section{Methods and Simulation Details}

From the solution of the BTE, the lattice thermal
conductivity is obtained as \cite{li_shengbte:_2014}
\begin{equation}
k_{l}^{\alpha\beta}=
\frac{1}{k_B T^2 \Omega N}
\sum_\lambda f_0\left(f_0+1\right)
\left(\hbar\omega_\lambda\right)^2
v_{\alpha,\lambda} F_{\beta,\lambda}
\label{equ.tc_liwu}
\end{equation}
where $\alpha$ and $\beta$ denote the $x$, $y$, or $z$ direction.
$k_B$, $T$, $\Omega$, and $N$ are Boltzmann constant, temperature, the
volume of the unit cell, and the number of $q$ points in the first
Brillouin zone respectively. The sum goes over phonon mode $\lambda$
that consists of both wave vector $q$ and phonon branch $\nu$. $f_0$ is
the equilibrium Bose-Einstein distribution function. $\hbar$ is the
reduced Planck constant. $\omega_\lambda$ is the phonon frequency and
$v_{\alpha,\lambda}$ is the phonon group velocity in $\alpha$
direction. The last term $F_{\beta,\lambda}$ is expressed in
Ref.~\citenum{li_shengbte:_2014} as
\begin{equation}
F_{\beta,\lambda}=\tau_{\lambda}
\left(v_{\beta,\lambda}+\Delta_\lambda\right)
\label{equ.term_F}
\end{equation}
where $\tau_{\lambda}$ is the phonon RTA lifetime. $\Delta_\lambda$
is a correction term that eliminates the inaccuracy of RTA by iteratively
solving BTE. When $\Delta_\lambda$ is fixed to be equal to zero, the RTA result for thermal conductivity is obtained. Equation~(\ref{equ.tc_liwu}) can be rearranged with the expression for volumetric phonon specific heat\cite{turney_predicting_2009} $c_{ph}$ and the RTA result for thermal conductivity becomes
\begin{equation}
k_{l}^{\alpha\beta}
=\frac{1}{N}\sum_\lambda
c_{ph,\lambda}
v_{\alpha,\lambda}
v_{\beta,\lambda}
\tau_{\lambda}
\label{equ.tc_cph_vg_tau}
\end{equation}
When the isotope scattering, boundary scattering and impurity scattering are ignored, the intrinsic three-phonon RTA lifetime $\tau_{\lambda}$ is computed as the inversion of the intrinsic scattering rate\cite{li_shengbte:_2014}
\begin{equation}
\tau_{\lambda}=\frac{1}{\Gamma_\lambda}=N\left(\sum_{\lambda^{'}\lambda^{''}}\Gamma^{+}_{\lambda\lambda^{'}\lambda^{''}}+\frac{1}{2}\sum_{\lambda^{'}\lambda^{''}}\Gamma^{-}_{\lambda\lambda^{'}\lambda^{''}}\right)^{-1}
\label{equ.tau}
\end{equation}
where $\lambda^{'}$ and $\lambda^{''}$ denote the second and third phonon mode that scatter with phonon mode $\lambda$. $\Gamma^{+}_{\lambda\lambda^{'}\lambda^{''}}$ and $\Gamma^{-}_{\lambda\lambda^{'}\lambda^{''}}$ are the intrinsic three-phonon scattering rates for absorption processes $\lambda+\lambda^{'}\to\lambda^{''}$ and emission processes $\lambda\to\lambda^{'}+\lambda^{''}$ respectively.
Both iterative method and RTA are used to predict thermal conductivity in our calculation. The method is discussed in detail in other literature.\cite{li_shengbte:_2014,omini_iterative_1995,turney_predicting_2009}

First-principles calculations were carried out using the
VASP package.\cite{kresse_efficiency_1996} In all our calculations, we
used the projector augmented-waves method\cite{blochl_projector_1994}
and the Perdew-Burke-Ernzerhof (PBE) exchange and
correlation.\cite{perdew_generalized_1996} A large energy cutoff of 400~eV was chosen. A vacuum spacing of 15~\AA~was used to prevent interactions between layers. The electronic stopping criteria was $10^{-8}$~eV. The hexagonal symmetry was enforced
during the geometry optimization. A hexagonal primitive unit cell
was first generated and optimized with 30$\times$30$\times$1 $k$-mesh for electronic integration, and then
a 5$\times$5$\times$1 supercell was built and re-optimized with a
6$\times$6$\times$1 $k$-mesh until the modulus of the force acting on each atom was less than $1.6\times 10^{-5}$ eV/\AA. For unstrained silicene, the external pressure in $xy$ plane was -0.02 kB after supercell optimization. The supercell was then used to compute the
force constants required for the phonon dispersion calculation. We
used the Phonopy package\cite{togo_first-principles_2008} to compute
and diagonalize the dynamical matrix and obtain the phonon dispersion
curve. The anharmonic force constants were extracted using the code
from ShengBTE package\cite{li_shengbte:_2014} called thirdorder.py. For
this calculation, up to the fourth nearest neighbors were considered. 
Thirdorder.py also applies the sum rules to the anharmonic
force constants. Finally, we use the ShengBTE package to compute the
thermal conductivity with the harmonic and anharmonic force constants.
A 101$\times$101$\times$1 $q$-mesh, that samples the first Brillouin
zone, and a thickness of 4.2~\AA, which is twice the van der Waals radius
of silicon, were considered. Isotope scattering with the natural isotopic distribution of silicon was
considered when solving the BTE, as implemented in
ShengBTE. We also tested all the strained cases without
isotope scattering and the results were quite similar.
For the strained structures we proceeded
in the same way. They were however optimized with a fixed strained
lattice constant. Strained structures were generated by stretching the
optimized unit cell by a certain percentage $\epsilon=(a-a_0)/a_0$, where  $a_0$ is the optimized lattice
constant of unstrained silicene and $a$ is that of strained silicene.

\section{Results and discussion}

\subsection{Structure}

\begin{figure}[h!]
\includegraphics[width=0.4\textwidth]{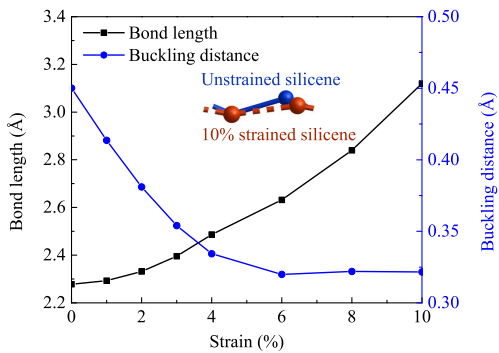}
\caption{Bond length and buckling distance as a function of strain. (Inset) Primitive unit cell structures for unstrained and 10\% strained silicene.}
\label{fig.structure}
\end{figure}

The optimized lattice constant of unstrained silicene is 3.87~\AA.
Figure \ref{fig.structure} shows the Si-Si bond length and buckling distance of
silicene as a function of strain.
The buckling distance first decreases significantly with increasing
strain from 0\% to 4\%, and then decreases slowly from 4\% to 6\%, and
finally stays almost unchanged from 6\% to 10\%. The small fluctuation
from 6\% to 10\% strain can be attributed to numerical uncertainty. On
the other hand, the Si-Si bond length keeps
increasing when strain becomes larger. This result is similar to
previous first-principles calculations.\cite{liu_strain-induced_2012}
It is known that $\pi$ bonding is weaker in silicene than that in graphene because of the longer bond distance. The $sp^2$ bonding will be dehybridized into $sp^3$-like bonding,\cite{cahangirov_two_2009,sahin_monolayer_2009} so silicene cannot have a complete planar structure as graphene,
even with large strain. We also perceive that the ratio of buckling
distance to bond length (nearest-neighbor distance) keeps decreasing with larger
strain, which means that the structure becomes more planar under larger
strain. The structures of unstrained silicene and silicene
under 10\% strain are shown in the inset of Fig. \ref{fig.structure}. It can be
clearly seen that strain will reduce the buckling distance and result
in longer atomic bond length.

\subsection{Dispersion}

\begin{figure}[h!]
\includegraphics[width=0.4\textwidth]{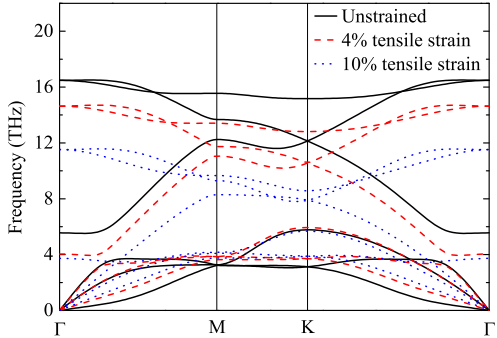}
\caption{Phonon dispersion curves of unstrained silicene and strained silicene under 4\% and 10\% tensile strain.}
\label{fig.dispersion}
\end{figure}

Figure \ref{fig.dispersion} shows the phonon dispersion curve in high-symmetry
directions. The phonon dispersion curve plays a crucial role in
computing the correct thermal
conductivity.\cite{paulatto_anharmonic_2013} Our result of phonon
dispersion curve for unstrained silicene is similar to other
first-principles calculations.
\cite{drummond_electrically_2012,scalise_vibrational_2012,sahin_monolayer_2009}
Since silicene has a small buckling, its structure does not have
reflectional symmetry\cite{lindsay_flexural_2010} across the $xy$ plane.
As a result, the vibrational pattern of flexural acoustic mode is not
purely out-of-plane, as demonstrated in our previous work.\cite{zhang_thermal_2014} Hereafter we denote this flexural acoustic
phonon branch as FA branch to avoid confusion with the purely
out-of-plane acoustic (ZA) branch in graphene. Another direct consequence
of the buckling is that the FA branch is not quadratic near the zone
center ($q\rightarrow 0$). Instead, it has a large linear
component\cite{gu_first-principles_2015} and therefore a
well-defined group velocity (Fig. \ref{fig.dispersion}). In addition, in the range of 3-6~THz, the longitudinal acoustic (LA)
branch and flexural optical (FO) branch have an avoided
crossing.\cite{delaire_giant_2011} This is again because the
LA and FO modes have the same
symmetry.\cite{dove_introduction_1993} This is different from graphene
where the out-of-plane optical (ZO)
and LA branch cross at about 25~THz
($\sim$834~\wn).\cite{yan_phonon_2008}

Comparing with the dispersion of unstrained silicene, the dispersion curves of optical phonon modes overall shift downward
when the applied tensile strain increases. This shift is mainly due to
the reduction of the material stiffness under tensile strain, which is
similar to other bulk and low dimensional
materials.\cite{li_strain_2010,picu_strain_2003} It is interesting to
note that the frequency gap at the avoided crossing is reduced when the
tensile strain increases. We believe this can be attributed to the fact
that the structure of strained silicene is becoming more planar when a
larger strain is applied, as we discussed earlier.

In order to quantify the acoustic phonon group velocity near the zone
center, we calculated the group velocities at $\Gamma$ point for three
acoustic modes. The group velocity of FA mode increases monotonically from 1075.7~m/s for unstrained
case to 3287.0~m/s for 10\% strain. This is related to the
increased stiffness in out-of-plane direction when we apply tensile
strain in the in-plane directions. The group velocity of LA mode
decreases monotonically from 9548.3~m/s for unstrained case to 7184.6~m/s for 10\% strain, which is due to the increase in the Si-Si bond length and
the weakened Si-Si atomic interaction in the in-plane directions.
For transverse acoustic (TA) mode, the group velocity first increases from 5629.7~m/s for unstrained case to
5804.4~m/s for 4\% strain, and then slightly decreases to 5235.7~m/s for 10\% strain. It should be noted that the strain
dependence of zone center phonon group velocity of silicene is quite
different from that of bulk silicon, in which the group velocity only
changes slightly under $\pm$3\% strain.\cite{parrish_origins_2014} The
strain dependence is quite similar to graphene, where the slope of ZA
branch increases with strain and LA branch decreases with
strain.\cite{fugallo_thermal_2014}

\subsection{Thermal conductivity}

\begin{figure}[h!]
\includegraphics[width=0.4\textwidth]{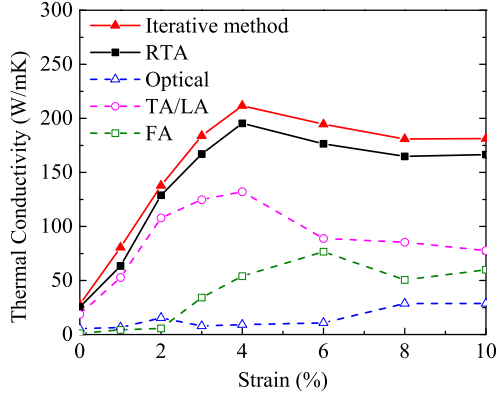}
\caption{Thermal conductivity of silicene as a function of strain
computed with iterative method and RTA. The dashed lines
represent the branch contribution of optical phonons, TA/LA phonons, and FA phonons to the total
thermal conductivity computed with RTA.}
\label{fig.TC}
\end{figure}

The thermal conductivity values of silicene at 300K under different tensile
strains are computed with both RTA and iterative method, as shown in
Fig. \ref{fig.TC}. The results of unstrained silicene are
25.5 and 27.9~W/mK for RTA and iterative method, respectively. This
is similar to the result of Gu \emph{et al.},
\cite{gu_first-principles_2015} in which the thermal conductivity of
unstrained silicene is predicted to be around 25~W/mK. We note that RTA and
iterative method give a similar trend in the strain dependence of
thermal conductivity of silicene. Both methods predict that
thermal conductivity first
increases significantly and then decreases slightly. The highest
thermal conductivity value (195.3~W/mK from RTA) appears
at 4\% strain and is about 7.5 times that of unstrained case. Such a
significant increase is quite anomalous. Usually the thermal
conductivity of a bulk material, such as
silicon,\cite{parrish_origins_2014} diamond,\cite{broido_thermal_2012}
and argon,\cite{chernatynskiy_thermal_2013}
would only slightly decrease under tensile strain. For graphene,
a similar method predicted that the thermal conductivity only slightly
increases with 4\% strain.\cite{fugallo_thermal_2014,kuang_unusual_2015} In some other
empirical molecular dynamics based calculations, it was even found that
the thermal conductivity of graphene decreases with tensile
strain.\cite{li_strain_2010} With all the strains we considered, the
maximum thermal conductivity occurs at 4\% strain, which is the turning
point where buckling distance of silicene does not decrease significantly
with strain any more. This implies that the thermal conductivity of silicene
have a strong correlation with the buckling distance. It should be noted that at 7.5\% strain silicene is predicted to become metallic,\cite{liu_strain-induced_2012} where electrons may also contribute to the total thermal conductivity. In our calculation, we considered the contribution of phonons to the thermal conductivity only. 

In the subsequent discussions, we choose the RTA result to explain this
anomalous behavior of silicene because phonon lifetime is well-defined
in RTA scheme. No extrapolation of thermal conductivity with respect to
$q$-mesh size is performed here because a denser $q$-mesh would produce slightly higher thermal conductivity and similar strain-dependence (we tested up to 201$\times$201$\times$1). It should be
pointed out that in Ref. \citenum{gu_first-principles_2015} it was
claimed that the thermal conductivity of silicene would diverge with
the sample size. Similar conclusions have been drawn in some other
literature for graphene. However, there has been strong debate on the
possible divergence of thermal conductivity of graphene. For example,
Fugallo \emph{et al.}\cite{fugallo_thermal_2014} argue that the
thermal conductivity of graphene will converge when the simulated
sampling size goes up to 1 mm. In their work, exact phonon BTE is
solved and first-principles calculations are employed to extract
harmonic and anharmonic interatomic force constants (IFCs). Barbarino \emph{et al.}
\cite{barbarino_intrinsic_2015} also reach the same conclusion with
approach-to-equilibrium molecular dynamics simulations for graphene
sample of 0.1 mm in size. With a finite $q$-mesh we sampled, we actually
exclude those extremely long wavelength acoustic phonon modes, which
are believed to be responsible for the possible divergence of the
thermal conductivity.\cite{bonini_acoustic_2012} For real applications
a finite sample has to be used, and the wrinkles and defects are
generally unavoidable, so the sample cannot have diverged thermal
conductivity. Since we are focusing on the strain effect of silicene,
we choose a finite $q$-mesh to avoid this issue.

To understand the anomalous strain dependence of thermal conductivity,
we first decompose the thermal conductivity contributions into
different phonon modes, and the results are also plotted in Fig.
\ref{fig.TC}. Different phonon modes are sorted by their frequencies.
The lowest branch is taken as the FA branch while the highest three
branches are taken as optical branches. The other two branches are
then TA or LA branch. This is a simple and commonly used method to
sort different phonon modes in the entire Brillouin zone. We have
carefully checked the symmetry of eigenvectors and find this algorithm
is reliable for almost all the phonons except at a few high symmetry
points. It is evident that the acoustic branches give the dominant
contribution over the full strain range, while contribution from
optical phonons is in the range of 4\% - 22\%. The percentage of
optical phonon contribution is similar to 
silicon nanowires.\cite{tian_importance_2011} Figure \ref{fig.TC} also shows that
LA and TA modes contribute more than half of the total thermal
conductivity and govern the trend of strain dependence of thermal
conductivity. Especially when the strain is smaller than 4\%, TA and
LA modes contribute more than 65\% of the total thermal conductivity.
This is quite different from graphene, for which it is believed that the
pure out-of-plane ZA mode has a major contribution to the total
thermal conductivity.
\cite{seol_two-dimensional_2010,lindsay_flexural_2010,lindsay_flexural_2011}
The thermal conductivity contributed by FA mode first increases with
strain up to 6\% and then slightly decreases. At zero strain, the
contribution from FA mode is only about 5\% while at 6\% strain the
contribution from FA mode increases up to around 45\%.

\subsection{Heat capacity, group velocity, and lifetime}

\begin{figure}[h!]
\includegraphics[width=0.4\textwidth]{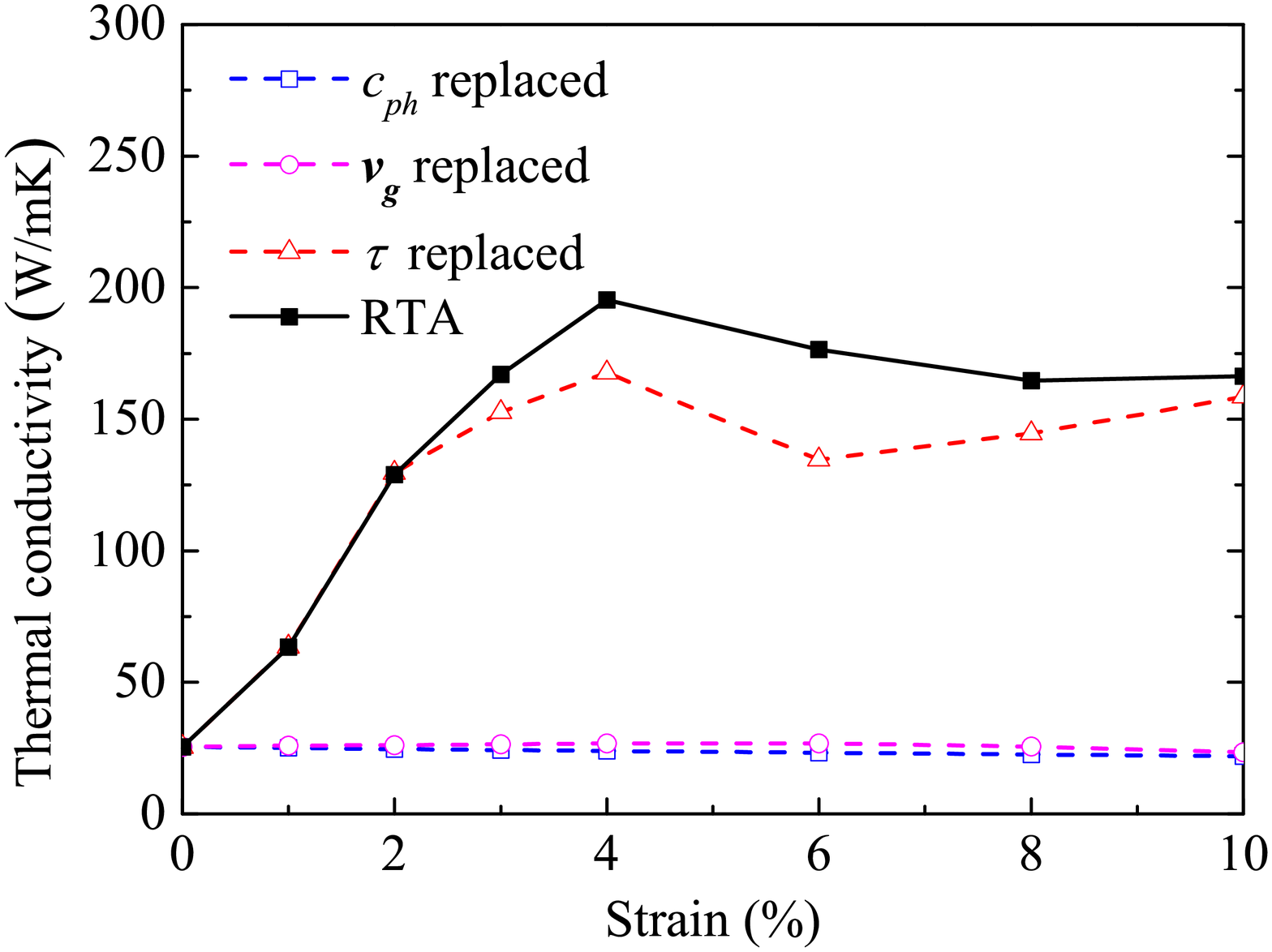}
\caption{Thermal conductivity as a function of strain computed with
RTA and cross-calculated thermal
conductivity with $c_{ph}$, $\boldsymbol{v_g}$ or $\tau$ replaced with strained values.}
\label{fig.CrossTC}
\end{figure}

From Eq. (\ref{equ.tc_cph_vg_tau}) we know that thermal conductivity is related to
volumetric phonon specific heat (heat capacity) $c_{ph}$, group velocity $\boldsymbol{v_g}$, and phonon lifetime $\tau$. In
order to find out the dominant factor for the anomalous strain
dependence of thermal conductivity, we substitute each of the three
terms for unstrained silicene with the value of strained silicene. For
example, when heat capacity is replaced, the thermal conductivity is
calculated as
\begin{equation}
k_{l}^{\alpha\beta,\epsilon}
=\frac{1}{N}\sum_\lambda
c_{ph,\lambda}^\epsilon
v_{\alpha,\lambda}^0
v_{\beta,\lambda}^0
\tau_{\lambda}^0
\label{equ.cross}
\end{equation}
where the superscript $\epsilon$ denotes the applied strain in
strained silicene and $0$ denotes unstrained silicene.
The results for three cases are plotted in Fig. \ref{fig.CrossTC}. We see that the calculated thermal conductivity changes
significantly when lifetime is replaced with the value of strained
silicene. At 4\% strain, the highest value is about 7 times that of
unstained case. This shows that the unusual strain
dependence of thermal conductivity is mainly due to the
change in phonon lifetime.
From 0\% to 6\% strain the thermal conductivity with lifetime replaced
changes dramatically, indicating that lifetime is the dominant factor
in this range of strain. From 6\% to 10\% strain the thermal
conductivity with lifetime replaced changes about 17.8\%, while the
thermal conductivity with group velocity or heat capacity replaced
changes -12.7\% and -5.7\% respectively. In this range of strain,
these changes are on the same order of magnitude. The three competing
factors balance in this range, so the change in the thermal
conductivity is small.

\subsection{Lifetime\label{sec.lifetime}}

\begin{figure}[h!]
\includegraphics[width=0.4\textwidth]{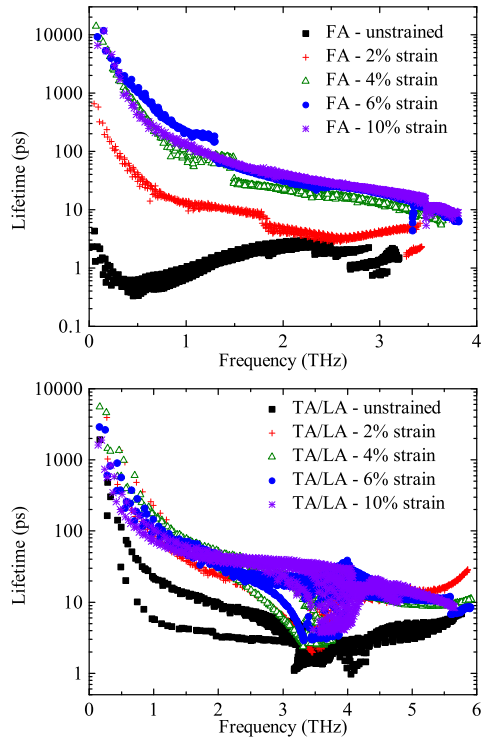}
\caption{(Top panel) Lifetime of FA phonons as a function
of frequency for 0\%, 2\%, 4\%, 6\%,
 and 10\% strain. (Bottom panel)
Lifetime of LA/TA phonons as a function of frequency for
0\%, 2\%, 4\%, 6\%, and 10\% strain.}
\label{fig.lifetime}
\end{figure}

To further quantify the lifetime variation under different strains, we
plotted the frequency dependent phonon lifetime in Fig. \ref{fig.lifetime}. Since acoustic phonons are the
dominant heat carriers in the thermal transport in silicene, we only
show the lifetimes for acoustic phonons. In addition, those negligible phonon modes whose aggregate contribution to thermal
conductivity is less than 0.1\% are excluded to reduce the amount of data points.
From Fig. \ref{fig.lifetime} it can be seen that, except for a few phonon
modes, the acoustic phonon lifetimes of strained silicene are
consistently significantly larger than that of unstrained case. The
top panel in Fig. \ref{fig.lifetime} indicates that the overall FA phonon
lifetimes keep increasing for strain from 0\% to 6\% but decrease a
little for strain from 6\% to 10\%. The major
heat carriers whose aggregate contribution to overall thermal conductivity is more than 50\% are those low-frequency acoustic phonons with frequencies under 2.8~THz. The bottom panel shows
that the lifetime of TA/LA phonons with frequencies lower than 2.8~THz would
overall increase when silicene is strained from 0\% to 4\%, and then
decrease afterwards. The transition from increased to slightly
decreased lifetime occurs in the range of 4-6\% strain for all the
important acoustic phonon modes, which is consistent with the
strain-dependent thermal conductivity. It should be noted that such a
significant change in phonon lifetime with tensile strain is quite
unusual. In Ref.\citenum{parrish_origins_2014}, the strain-dependent
phonon lifetime for solid argon and silicon was calculated using a
similar approach. The phonon lifetimes of both materials show quite small
strain dependence (changes are within 300\%). Here for silicene, the
phonon lifetime of the majority of low frequency FA phonons increases
by two to three orders of magnitude with strain and the variation of
TA/LA lifetime is also one to two orders of magnitude.

%This is evidently related to the
%significant change of the dispersion curves of acoustic phonons. We
%also anticipate that the reduction of the phonon scattering of FA
%modes with other modes is associated with the change of the atomic
%structure (especially the decrease in the buckling distance). This is
%in line with the discussion of phonon lifetime in graphene\cite{lindsay_flexural_2010}: it was
%shown that the reflectional symmetry (zero buckling distance) will
%eventually forbid the phonon scattering process with odd number of ZA
%mode, which results in a large ZA phonon lifetime.

\subsection{Weighted phase space and scattering channel}
To understand such a dramatic change of acoustic phonon lifetime, we 
also calculated the phase space defined in 
Ref.~\citenum{lindsay_three-phonon_2008} and the ``weighted phase space" 
defined in Ref.~\citenum{li_ultralow_2015} (results not shown). The 
weighted phase space is an expression of frequency-containing factors 
that quantifies the phonon scattering probability for a particular 
dispersion curve. The weighted phase space was used to successfully 
explain the ultralow thermal conductivity of filled skutterudite 
YbFe$_{4}$Sb$_{12}$.\cite{li_ultralow_2015} However, unlike 
YbFe$_{4}$Sb$_{12}$ whose low thermal conductivity is mainly attributed 
to the allowed phase space for scattering, the variation of weighted 
phase space of silicene under different strains cannot fully explain the 
large variation of phonon lifetime. We also calculated thermal 
conductivity of using the harmonic force constants of 4\% strain and 
anharmonic force constants with 0\% strain, the result is 
only about twice the unstrained thermal conductivity. Together with the 
calculation result of weighted phase space, we find that the variation 
of neither harmonic nor anharmonic force constants alone can fully 
account for the variation of phonon lifetime.

\begin{widetext}
\begin{figure*}[h!]
\includegraphics[width=1.0\textwidth]{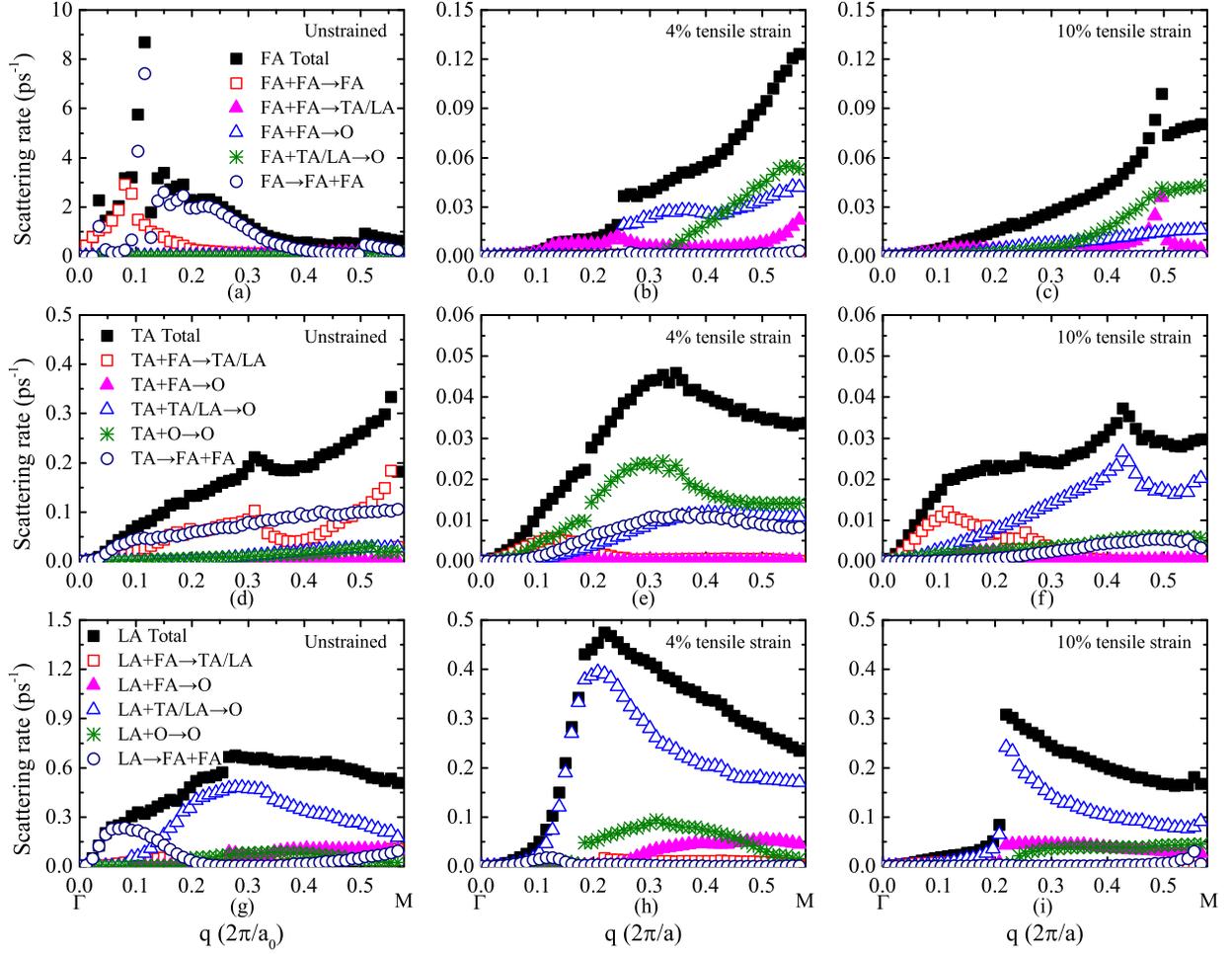}
\caption{Scattering rates of acoustic phonons from $\Gamma$ to M:
FA phonons for (a) unstrained silicene, (b) 4\% strained silicene, (c)  10\% strained silicene. TA phonons for (d) unstrained silicene, (e) 4\% strained silicene, (f) 10\% strained silicene. LA phonons for (g) unstrained silicene, (h) 4\% strained silicene, (i) 10\% strained silicene. (Please note the difference in the scales for scattering rates in the subfigures.)}
\label{fig.channel}
\end{figure*}
\end{widetext}

We further investigate the different scattering channels to quantify the 
importance of different phonon modes in the scattering processes. Figure 
\ref{fig.channel} shows the scattering rates of acoustic phonons along 
$\Gamma$ to M direction, and only major scattering channels that has a 
large contribution to overall scattering rate are included. Scattering rates for 
emission processes are multiplied by 1/2 to avoid counting the same 
process twice. Note that along $\Gamma$ to M direction the FA, TA, and 
LA branches can be easily separated, so the scattering channels for the 
three branches are plotted separately in Fig. \ref{fig.channel}.

Figure \ref{fig.channel} (a,b,c) show the scattering rates of different scattering channels of FA phonons for unstrained (0\%), 4\%, and 10\% strained silicene respectively. Note that the legends are the same for these three subfigures while the scales for the $y$-axis are not the same. It can be seen that the total scattering rates of FA phonons decrease orders of magnitude  from 0\% to 4\% but decrease only a little from 4\% to 10\%. In unstrained silicene, dominant scattering channels are the scattering among the FA modes (i.e., FA+FA$\to$ FA and FA$\to$FA+FA processes). However for strained silicene, either 4\% or 10\%, the dominant scattering channels become FA+FA$\to$TA/LA, FA+FA$\to$O and FA+TA/LA$\to$O, where O indicates optical phonons. Our data also show that scattering rates of FA+TA/LA$\to$O processes also decrease from 0\% to 4\% strain. We attribute the decrease of scattering rates for processes involving odd number of FA phonons to the change of buckling distance. For graphene, because of the reflectional symmetry of the structure, the phonon scattering process involving odd number of ZA phonons is not allowed, leading to a very small scattering rate of ZA mode \cite{lindsay_flexural_2010}. Our observation of FA mode is in line with their discussion on graphene: when the strain is larger, the silicene structure becomes more planar, so the scattering processes involving odd number of FA modes (especially the scattering process involving 3 FA modes) decreases.

As we noted before, the dramatic change of thermal conductivity from 0\% to 4\% is mainly due to the in-plane modes (TA and LA). Therefore, we need to further check the scattering channel of TA and LA modes. Figure \ref{fig.channel} (d,e,f) plot the scattering rates of TA phonons for 0\%, 4\%, and 10\% strained silicene respectively. Similar to the trend of FA phonons, the total scattering rates reduces significantly from 0\% to 4\% strain. From 4\% to 10\% strain, scattering rates still decrease slightly in the high frequency region but increase in low frequency. For unstrained silicene, dominant scattering channels are TA+FA$\to$TA/LA and TA$\to$FA+FA. For silicene under 4\% tensile strain, dominant scattering channels are TA+O$\to$O, TA+TA/LA$\to$O and TA$\to$FA+FA. For silicene under 10\% strain, dominant scattering channels become TA+TA/LA$\to$O and TA+FA$\to$TA/LA. The variation of scattering rates for different scattering channels for TA phonon modes are the most complicated. The scattering rates of the common dominant process TA$\to$FA+FA become smaller with larger strain. For TA+TA/LA$\to$O process, the scattering rates increase. For TA+O$\to$O, the scattering rates first become larger from 0\% to 4\%, and then become smaller from 4\% to 10\%. For TA+FA$\to$TA/LA, the trend is opposite to previous one: scattering rates first become smaller then become larger. Overall, we find that with larger strain the scattering with FA phonon becomes weaker while the scattering with optical phonon becomes stronger. These competing mechanisms result in the change of the total scattering rates.

Figure \ref{fig.channel} (g,h,i) at the bottom are the scattering rates of LA phonons for 0\%, 4\%, and 10\% strained silicene respectively. The total scattering rates of LA phonons decrease monotonically from 0\% to 4\% and then to 10\% strain, but we also note that the total scattering rate of LA phonon does not change as much as FA and TA modes. For different strains LA+TA/LA$\to$O is the dominant scattering channel for all the three cases, presumably due to the relatively higher frequency of LA mode. At the low frequency regime of the unstrained case the LA$\to$FA+FA is dominate, but at larger strain this channel is becoming less important.

From the analysis of the scattering channel, we have a few observations. First, the scattering process among FA mode is significantly reduced with larger tensile strain, which is due to the reduced buckling distance and more planar structure. This could explain the significant enhancement of phonon lifetime for FA modes as seen in Fig. \ref{fig.lifetime}. We should note that this is not the major reason for the enhanced thermal conductivity from 0\% to 4\% strain, because FA modes have relatively small contributions to thermal conductivity in this range. Second, the scattering rates of TA modes decrease significantly from 0\% to 4\% strain, mainly due to the reduced scattering with FA modes. This is the major factor that the thermal conductivity of silicene increases in this range. Third, the LA phonon scattering rates do not change significantly under different strains and thus are not responsible much to the large tunability of thermal conductivity. Lastly, we should note that the above analysis is based on the $\Gamma$ to M region of the Brillouin zone. To distinguish TA and LA phonon is not always possible for any $q$ point, so it is not safe to conclude that the enhanced thermal conductivity is mainly due to TA modes. We rather believe that the enhanced thermal conductivity is mainly due to the enhanced lifetime of both LA and TA modes because they scatter less with FA mode when strain is applied.

\section{Summary}
In conclusion, we performed first-principles calculations to predict the
lattice thermal conductivity of silicene under strain. Phonon BTE is
solved both in the RTA scheme and iteratively in our prediction. Both methods yield a similar trend in the change of thermal
conductivity with respect to tensile strain. It is shown that within
10\% tensile strain the thermal conductivity of silicene first increases
dramatically and then decreases slightly. The maximum thermal
conductivity was found when 4\% tensile strain was applied, and the
value was about 7.5 times that of the unstrained case. Such a dramatic
change is quite unusual for solid materials, and could be used as a
thermal switch together with thermal diodes to build thermal circuits.
This trend is mainly due to the strain-dependent phonon lifetime, which is related to the variations of both harmonic and anharmonic force constants under strain. FA phonon lifetimes increase significantly under tensile strain because the structure becomes more planar. This leads to a large increase of their contribution to overall thermal conductivity, but is not the major reason for the significant change of overall thermal conductivity within 4\% strain. The significant enhancement of thermal conductivity from 0\% to 4\% strain is mainly due to the reduced scattering of TA and LA phonons with FA phonons. Our result suggests that other 2D materials with intrinsic buckling may have similar strain dependence of thermal conductivity, which is left for further investigation.

\begin{acknowledgements}
This work was supported by the National Natural Science Foundation of
China (Grant No.~51306111) and Shanghai Municipal Natural Science
Foundation (Grant No.~13ZR1456000). M.H. acknowledges the support
by the Deutsche Forschungsgemeinschaft (DFG) (project number:
HU~2269/2-1). Simulations were performed with computing resources
granted by HPC ($\pi$) from Shanghai Jiao Tong University. The authors
thank Dr. Wu Li for answering questions regarding ShengBTE.
\end{acknowledgements}

\end{document}